\documentstyle[seceq,preprint,epsf,wrapft]{ptptex}

\newcommand{\nn}{\nonumber\\}

\newcommand{\abs}[1]{\left| #1 \right|}
\newcommand{\bra}[1]{\left< #1 \right|}
\newcommand{\ket}[1]{\left| #1 \right>}

\newcommand{\bidx}[1]{{\textstyle {\atop #1}}\!\!}
\def\VEV#1{\left<#1\right>}
\def\v(#1){\bra{v(#1)}}
\def\calH{{\cal H}}
\def\calO{{\cal O}}
\def\m1{{-1}}
\def\ketR(#1#2){\left|R(#1,#2)\right\rangle}
\def\braR(#1#2){\left<R(#1,#2)\right|}
\def\Tr{\mathop{\rm sTr}}
\preprintnumber[3cm]{
KUNS-1509\\ HE(TH)~98/7\\ UT-813\\ hep-th/9805119}

\title{
\hspace{-0.5cm}
On the Generalized Gluing and Resmoothing Theorem
}

\author{
Tsuguhiko {\sc Asakawa},\footnote{E-mail: asakawa@gauge.scphys.kyoto-u.ac.jp}
 Taichiro {\sc Kugo}\footnote{E-mail:
kugo@gauge.scphys.kyoto-u.ac.jp}
and Tomohiko {\sc Takahashi}$^{\dagger ,}$\footnote{
JSPS Research Fellow. E-mail: tomo@hep-th.phys.s.u-tokyo.ac.jp}
}

\inst{
Department of Physics, Kyoto University, Kyoto 606-01
\\
$^\dagger $Department of Physics,
University of Tokyo, Tokyo 113
}

\recdate{
\today
}

\abst{
The generalized gluing and resmoothing theorem originally proved by 
LeClair, Peskin and Preitschopf, gives a powerful formula for the fused 
vertex obtained by contracting any two vertices in string field theories. 
Although the theorem is naturally expected to hold for the vertices at 
any loop level, the original proof was restricted to the vertices at 
tree level. Here we present a simplified proof for the tree level 
theorem and then prove explicitly the extended version at one-loop 
level. We also find that a non-trivial sign factor, which depends on the 
string states to be contracted, appears in the theorem. This sign factor 
turns out to be essential for reproducing correctly the conformal field 
theory correlation function on the torus.}

\begin{document}

\maketitle

\section{Introduction}

The basic ingredients of string field theories (SFT) are the 
vertices. We need to combine (or fuse) the various vertices in many 
situations, e.g.,
in showing the gauge invariance of the SFT action and in computing string 
scattering amplitudes in perturbation theory. 
LeClair, Peskin and Preitschopf\cite{rf:LPP1} (LPP) have developed a powerful
method for defining general multistring vertices by conformally mapping the
unit disks of participating strings into a complex plane and by using 
the correlation functions of conformal field theory\cite{rf:BPZ,rf:FMS} 
(CFT) in the plane. 
They then proved a theorem, which they called ``Generalized Gluing and 
Resmoothing Theorem" (GGRT),\cite{rf:LPP2} giving a general formula 
for the fused 
vertex obtained by a contraction of two vertices. They showed that the 
fused vertex is just equals their multistring vertex corresponding to 
the conformal mappings induced by gluing the two world sheets into one. 

The point here is that the equality holds with {\it weight one} 
if the conformal anomaly (i.e., central charge $c$) is zero.
Their proof is very thorough and even pedagogical. It is, however, 
a bit complicated and they did not take much care about the sign 
of the equality. The relevant sign factor is not a mere phase but an 
operator which changes sign depending on the string states to be 
contracted, and become very important, for example, for the 
cancellations of divergences between two graphs\cite{rf:KT} 
and also in the proof of
the gauge invariance of the SFT action.\cite{rf:AKT} \ \ 
In the context of this paper also,
the sign factor in the theorem at tree level turns out to be essential for 
reproducing correctly the conformal field theory correlation function 
on the torus.

Here in this paper we first present a much simplified proof for the LPP 
GGRT, and determine the sign of the equality carefully. Our proof is 
inspired by the sewing method of two conformal field theories defined on two 
Riemann surfaces, which has been given, in particular, 
by Sonoda.\cite{rf:sonoda} 
Actually, the GGRT by LPP is a SFT version of this general 
way of sewing two CFT's. The gluing and the sewing are essentially the 
same and are just the insertion of the complete set of states. So, 
although the original GGRT by LPP is restricted to the vertices at tree 
level, it is naturally expected to hold at any loop level. Nevertheless,
the SFT version is not so trivial. This is because the gluing in SFT 
must be performed by contracting two strings, one each from the two 
vertices. On the other hand the sewing of two CFT's is performed by 
excising two holes freely, one each on the two Riemann surfaces. To do 
the same thing in the SFT case and to make contact with the definition 
of the vertices, one needs to map the string world sheets back and forth. 
These mappings give non-trivial conformal transformations on the operators, 
which must be traced neatly. We perform this procedure and
prove explicitly an extended version of the GGRT at one loop level.
It may be interesting to note that, as a byproduct, 
the formula for the CFT correlation function on the torus\cite{rf:FGST} 
\begin{equation}
\left< \, \calO_1\,\calO_2\,\cdots\, \right>_{{\rm torus}\ \tau}
={\rm Tr}\,[(\m1)^{N_{\rm FP}} \,q^{2L_0}\,\calO_1\,\calO_2\,\cdots\, ]
\end{equation}
is automatically {\it derived} by this procedure (up to an overall factor 
convention), where $q=e^{i\pi\tau}$ and $N_{\rm FP}$ is the ghost number 
operator. The factor $(\m1)^{N_{\rm FP}}$ in this expression comes from 
the operator sign factor in the theorem at tree level mentioned above.

This paper is organized as follows. First, in \S 2, we briefly review the 
the definition of the vertices and the GGRT given by LPP. In \S 3, 
after making some remarks on the ambiguity present when $c\not=0$ 
in defining conformal transformation operators $U_f$ corresponding to 
the mappings $f(z)$, 
we present two propositions to clarify when $U_f$ leaves inert the 
SL(2;C) bra and ket vacua, and then, gives a simplified proof for the GGRT 
of LPP. The extension of GGRT to one-loop level is given and proved in
\S 4. 

For simplicity of presentation, we assume henceforth that the strings are 
all bosonic open strings, so that the relevant conformal fields
$\phi(z)$ are string coordinates $\partial X^\mu(z)$, and reparameterization ghost 
$c(z)$ and anti-ghost $b(z)$, possessing dimensions $d=1,\ -1$ and $2$,
respectively.  Closed string can be treated similarly since it is 
more or less equivalent with a pair of open strings. 

\section{GGRT at tree level}
First of all, let us recall LPP's definition of the tree level vertex 
which refers to the conformal field theory in the complex plane (two 
dimensional manifold $M$ which is topologically equal to 
$S^2$):\cite{rf:LPP1,rf:KS,rf:AGMV}
\begin{equation}
\v(n,\cdots,2,1) \ket{A_1}_1
\ket{A_2}_2\cdots\ket{A_n}_n
\equiv \bigl\langle \,h_1[\calO_{A_1}] h_2[\calO_{A_2}]
\cdots h_n[\calO_{A_n}]\, \bigr\rangle_M\ .
\label{eq:LPPdef}
\end{equation}
Here $\v(n,\cdots,2,1)$ is the $n$-point LPP vertex, which is defined as 
a bra state in the product space $\otimes_{i=1}^n \calH_i$ of $n$ 
string Fock spaces $\calH_i$, and each string state 
$\ket{A_i}_i\in \calH_{i}$ are given in the form
\begin{equation}
\ket{A}_{i} = \calO_A\ket{0}_i\,,
\end{equation}
where $\calO_A$ is an operator creating the state $A$ of string $i$ from 
the SL(2;C) invariant vacuum $\ket{0}_i$ in $\calH_i$; 
for instance, the tachyon state of momentum $p$ is given by the vertex 
operator $\calO(w)=c(w)\exp(ip\cdot X(w))$ at $w=0$, and the ladder 
operators $\phi_n=\{\alpha_n,\ c_n,\ b_n\}$ are given by the contour integration
$\oint (dw/2\pi i)w^{n+d-1}\phi(w)$ encircling the origin. 

The meaning of the right-hand side of Eq.~(\ref{eq:LPPdef}) is 
as follows: any vertex $\v(n,\cdots,2,1)$ is defined by specifying how
the participating strings $i$ are glued each other. We can regard 
each string world sheet from infinite past ($\tau_i=-\infty$) to the 
interaction time ($\tau_i=0$) as a 
unit disk $\abs{w_i}\leq1$ with $w_i=\exp(\tau_i+i\sigma_i)$, and the world 
sheet formed by gluing those string sheets as a complex $z$-plane, 
which we call $M$ ($\sim S^2$), for tree level vertex case. So this gluing 
can be simply specified by giving conformal mappings $h_{i}(w_i)$ of 
each string $w_{i}$ plane into the complex $z$ plane $M$, which is 
analytic and invertible inside the each unit circle $\abs{w_i}=1$. 
Generally, any conformal mapping $f:\ w\ \rightarrow\ z=f(w)$ of $w$-plane to 
$z$-plane also defines a mapping of operators $\calO$ in the $w$-plane 
to operators $f[\calO]$ in the $z$-plane:
\begin{equation}
f[\calO] \equiv U_f\, \calO\ U^{-1}_f\,.
\label{eq:Uf}
\end{equation}
If the operator $\calO$ is a primary conformal field $\phi(w)$ of dimension 
$d_\phi$, this mapping is defined to be
\begin{equation}
f\bigl[\phi(w)\bigr] = \left({df(w)\over dw}\right)^{d_\phi}\phi\bigl(f(w)\bigr).
\label{eq:OpTrf}
\end{equation}
The operator representation $U_f$ in Eq.~(\ref{eq:Uf}) of the conformal 
mapping $f$ is uniquely determined by this transformation law 
(\ref{eq:OpTrf}) of the primary fields {\it up to a multiplicative 
constant}. Since the Fourier components of the energy momentum tensor 
$T(z)$, $L_n\equiv\oint (dz/2\pi i) z^{n+1}T(z)$, generate infinitesimal 
conformal transformations, the operator $U_f$ for the finite 
transformation $f$ can be given in the form
\begin{equation}
U_f = \exp\bigl(\sum_n v_{-n}L_n\bigr)
\label{eq:Ufcanonical}
\end{equation} 
with certain parameters $v_n$. ($v(z)\equiv\sum_n v_{-n}z^{n+1}$ and $f(z)$ 
are related by $f(z)=e^{v(z)\partial_z}z$.\cite{rf:LPP2}) \ \ 
We should keep in mind, however, that this parameterization 
(\ref{eq:Ufcanonical}) for $U_f$ is 
not unique and that the very definition of $U_f$ by 
Eq.~(\ref{eq:Uf}) has an ambiguity of multiplicative constant. 
We shall come back to this problem later in the next section. 

Now the meaning of the right-hand side of Eq.~(\ref{eq:LPPdef}) will be 
clear: it gives a correlation function of the mapped operators 
$h_i[\calO_{A_i}]$ of the conformal field theory in the $z$ plane. 
Note the crucial conceptual difference between both sides of 
the defining equation (\ref{eq:LPPdef}) of LPP vertex; the left-hand side 
is an inner product in the product space $\otimes_{i=1}^n \calH_i$ of $n$ 
string Fock spaces $\calH_i$, while the right-hand side is 
a correlation function of {\it a single string} conformal field theory in 
$z$ plane.

We need a little more preparation to state the GGRT. Let us introduce
bra and ket reflectors $\braR(12)$ and $\ketR(12)$ which convert 
ket string states $\ket{A}$ to bra states $\bra{A}$, and vice versa:
\begin{equation}
\braR(12)\ket{A}_2 = \bidx{1}\bra{A}, \qquad 
\bidx{2}\bra{A}\ketR(21) = \ket{A}_1\,.
\end{equation}
The reflectors $\braR(12)$ and $\ketR(12)$ are just the metric $g_{IJ}$ 
and $g^{IJ}$, respectively, if we use notation $\ket{A}\equiv A^I$ and 
$\bra{A}\equiv A_I$. So they can be defined by giving an inner product 
in the string Fock space $\calH$. A natural inner product\cite{rf:BPZ} 
is defined by using the inversion $I(z)=-1/z$ as follows:
\begin{eqnarray}
\left<A|B\right> &=& \braR(12)\ket{A}_2\ket{B}_1\nn
&=& \left<\,I[\calO_A]\,\calO_B\,\right>\,.
\end{eqnarray}
It is easy to find an explicit oscillator expression for the reflectors,
as can be found, e.g., in Refs.~\citen{rf:LPP2} and \citen{rf:HIKKO1}. 
We here need not that explicit expression but the following formal one. 
Let $\{\, \ket{\alpha}\, \}$ be a complete set of the ket string states 
and $\{\, \bra{\tilde \alpha}\, \}$ be its orthonormal dual under this 
inner product; i.e., $\langle\tilde \beta|\alpha\rangle= 
\langle\,I[\calO_{\tilde\beta}]\,\calO_\alpha\,\rangle= 
\delta^\alpha_\beta$. Then we have a completeness relation:
\begin{equation}
\sum_\alpha \ket{\alpha}\bra{\tilde\alpha} = 
\sum_\alpha \calO_\alpha\ket{0}\bra{0}I[\calO_{\tilde \alpha}] = {\bf 1}  
\qquad {\rm in}\quad \calH,
\label{eq:completeness}
\end{equation}
where $\calO_\alpha$ and $\calO_{\tilde \alpha}$ are operators creating
the states $\ket{\alpha}$ and $\ket{\tilde\alpha}$, respectively. It is now clear 
that the reflectors have the following formal expressions:
\begin{eqnarray}
\braR(12) &=& \sum_\alpha \bidx{1}\bra{\alpha}\bidx{2}\bra{\tilde\alpha} = 
\sum_\alpha \bidx{1}\bra{0}I[\calO_\alpha]
\ \bidx{2}\bra{0}I[\calO_{\tilde \alpha}]\,, \nn
\ketR(12) &=& \sum_\alpha \ket{\alpha}_1\ket{\tilde\alpha}_2 = 
\sum_\alpha \calO_\alpha\ket{0}_1\ \calO_{\tilde \alpha}\ket{0}_2\,.
\label{eq:reflector}
\end{eqnarray}

Now we define the contraction (or fusion) of two vertices 
appearing in the GGRT. 
Let $\v(C,\{A_i\})$ be an 
$(n{+}1)$-point LPP vertex for the strings $A_i$ ($i=1,2,\cdots,n$) and $C$ 
defined by conformal mappings $h_{A_i}$ and $h_C$:
\begin{equation}
\v(C,\{A_i\}) \prod_{i=1}^n \ket{A_i}_{A_i}
\ket{C}_C
\equiv \bigl\langle \,\prod_{i=1}^n h_{A_i}[\calO_{A_i}] \, h_C[\calO_{C}]
\, \bigr\rangle_M\ .
\label{eq:vertexA}
\end{equation}
And let $\v(D,\{B_j\})$ be another $(m{+}1)$-point LPP vertex defined 
similarly:
\begin{equation}
\v(D,\{B_j\}) \prod_{j=1}^m \ket{B_j}_{B_j}
\ket{D}_D
\equiv \bigl\langle \,\prod_{j=1}^m h_{B_j}[\calO_{B_j}] \, h_D[\calO_{D}]
\, \bigr\rangle_N\ .
\label{eq:vertexB}
\end{equation}
Note that we have called the $z$-planes for the two cases $M$ and $N$, for 
distinction, although they are both topologically $\sim S^2$. 
Then we can define a fused vertex $\v(\{B_j\},\{A_i\})_{\rm fused}$ of 
these two vertices by gluing the strings $C$ and $D$ in each by the help
of the ket reflector $\ketR(CD)$: 
\begin{equation}
\v(\{B_j\},\{A_i\})_{\rm fused} \equiv \v(D,\{B_j\}) \v(C,\{A_i\}) \ketR(CD)\,. 
\label{eq:fusedVertex}
\end{equation}
Intuitively, the fusion gives a new Riemann surface which is formed by 
cutting out the images of the unit disks of strings $C$ and $D$ in $M$ 
and $N$, respectively, and then gluing smoothly the rest pieces of $M$ 
and $N$ together. This Riemann surface again becomes a complex plane, a 
manifold $M\infty N$ topologically $\sim S^2$. This gluing also induces 
conformal mappings $\hat h_{A_i}(w_i)$ and $\hat h_{B_j}(w_j)$ of the 
unit disks $\abs{w_i}\leq1$ and $\abs{w_j}\leq1$ of $n+m$ strings 
$\{A_i\},\{B_j\}$ into the plane $M\infty N$, which are again analytic and 
invertible inside each disk. 

Now we can state the GGRT, which was first proved by LPP\cite{rf:LPP2} 
aside from the sign factor:

\medskip
{\bf Theorem}[LPP] \quad {\it 
Let $\v(\{B_j\},\{A_i\})$ be the LPP vertex defined by this set of mappings, 
\begin{equation}
\v(\{B_j\},\{A_i\})
\prod_{i=1}^n \ket{A_i}_{A_i}
\prod_{j=1}^m \ket{B_j}_{B_j}
= \bigl\langle \,\prod_{i=1}^n \hat h_{A_i}[\calO_{A_i}] \, 
\prod_{j=1}^m \hat h_{B_j}[\calO_{B_j}] \, \bigr\rangle_{M\infty N}\ .
\label{eq:GGRTvertex}
\end{equation}
Then, if the conformal anomaly is zero, the fused vertex 
(\ref{eq:fusedVertex}) is just equals this LPP vertex up to a sign 
factor $\epsilon(A)$:
\begin{equation}
\v(D,\{B_j\}) \v(C,\{A_i\}) \ketR(CD) = 
\epsilon(A)\,\v(\{B_j\},\{A_i\})\,.
\label{eq:GGRT}
\end{equation}
The sign factor $\epsilon(A)$ is an operator 
\begin{equation}
\epsilon(A) = (\m1)^{\sum_i\abs{A_i}}
\label{eq:sign}
\end{equation}
which changes sign depending on the states 
$\ket{A_i}=\calO_{A_i}\ket{0}$ 
to be contracted, where ${\abs{A_i}}$ denotes statistics index 
defined to be $0$ $(1)$ when the operator $\calO_{A_i}$ is bosonic 
$($fermionic$)$. If the conformal 
anomaly is present, the equality (\ref{eq:GGRT}) is violated by a 
multiplicative c-number factor which depends non-trivially on the mappings 
$g$, $h_D$ and $h_C$}.
\medskip

\begin{figure}[htb]
   \epsfysize= 13cm
   \centerline{\epsfbox{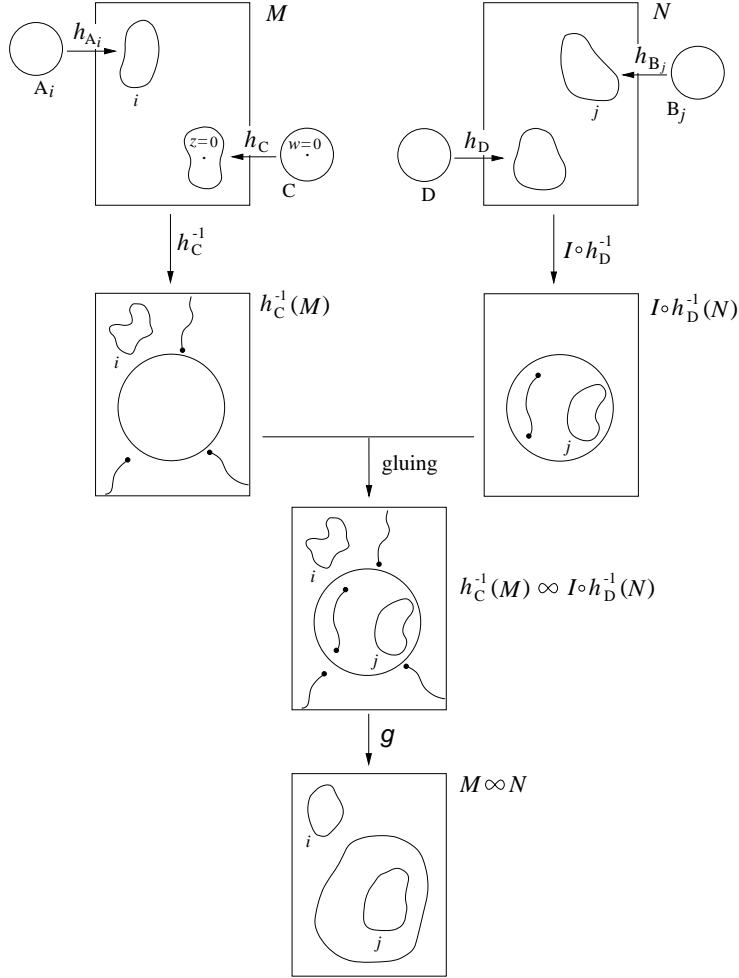}}
 \caption{Gluing and subsequent smoothing in the contraction of two 
          vertices.}
 \label{fig:1}
\end{figure}

LPP analyzed the above gluing procedure in Eq.~(\ref{eq:fusedVertex}) 
more carefully as shown in Fig.~\ref{fig:1}. First, 
the complex planes $M$ and $N$ defining the vertices $\v(C,\{A_i\})$ 
and $\v(D,\{B_j\})$ are mapped by $h_C^\m1$ and $I\circ h_D^\m1$ 
so that 
the exterior region of string $C$ in $M$ is mapped to the region outside 
a unit circle and the exterior region of string $D$ in $N$ to the 
inside of a unit circle, respectively. Then the region outside the 
unit circle in the plane $h_C^\m1(M)$ and the region inside the unit 
circle in the plane $I\circ h^\m1_D(N)$ are glued smoothly as they stand. 
But, unless the mappings $h_C$ and $h_D$ are $SL(2;C)$ transformations, 
neither they nor their inverses will be one-to-one mappings outside the 
unit circles. So the glued surface $h_C^\m1(M)\infty\,I{\circ}h^\m1_D(N)$ 
will generally possess branch-cut singularities. Since the covering 
surface nevertheless has a topology of $S^2$, there exists a mapping $g$ 
which carries the surface into the plane $M\infty N$, smoothing out the 
branch cuts. Therefore the conformal mappings $\hat h_{A_i}$ and $\hat 
h_{B_j}$ of the strings $\{A_i\},\{B_j\}$ into the plane $M\infty N$, 
mentioned in the Theorem, can thus be identified with
\begin{equation}
\hat h_{A_i} = g\circ h_C^\m1\circ h_{A_i}\,,
\qquad 
\hat h_{B_j} = g\circ I\circ h_D^\m1\circ h_{B_j}\,.
\label{eq:mappingrel}
\end{equation}
The last step mapping $g$ corresponds to a resmoothing procedure, 
explaining the name `Gluing and Resmoothing Theorem'. 

\section{A simple proof for the tree level GGRT}

We give in this section a proof for the GGRT at tree level. This proof 
is much simpler than the original one by LPP and, therefore, makes it 
easy to trace correctly the appearing sign factors. 

As promised, we first discuss the parameterization forms 
for the conformal transformation operator $U_f$ introduced in 
Eq.~(\ref{eq:Uf}), and its ambiguity of multiplicative constant which 
exists if {\it the conformal anomaly (central charge $c$) is nonzero}. 
As in the usual Lie group, there are a variety of ways of representing the 
group elements $U_f$ in terms of the Virasoro generators $L_n$.  
We call the parameterization form 
\begin{equation}
U_f=\exp\bigl(\sum_n v_{-n}L_n\bigr)
\end{equation}
already cited in Eq.~(\ref{eq:Ufcanonical}) `canonical form', 
which is most commonly used in the Lie group theory. 
Another useful parameterization form which we refer to as
`normal ordered' form, is given by
\begin{equation}
U_f = \exp\bigl(\sum_{n\geq2} v_{n}L_{-n}\bigr)
\exp\bigl(\sum_{k=0,\pm1} v_{-k}L_k\bigr) 
\exp\bigl(\sum_{m\geq2} v_{-m}L_m\bigr)\,. 
\label{eq:Ufstandard}
\end{equation}
Note that the middle factor $\exp\bigl(\sum_{k=0,\pm1} 
v_{-k}L_k\bigr)$ is the element belonging to the $SL(2;C)$ subgroup. 
Of course we can convert various parameterization forms from one to 
another by using the commutation relations of $L_n$. But the point is 
that, if the conformal anomaly is nonzero, there appears a non-trivial 
multiplicative c-number factor in front in this rewriting; for instance,
we have a relation like
\begin{equation}
\exp\bigl(\sum_{n\geq2} v_{n}L_{-n}\bigr)
\exp\bigl(\sum_{k=0,\pm1} v_{-k}L_k\bigr) 
\exp\bigl(\sum_{m\geq2} v_{-m}L_m\bigr) 
= e^a\,
\exp\bigl(\sum_n v'_{-n}L_n\bigr)\,.
\end{equation}
The front c-number factor $e^a$ depends on the central charge $c$
(the exponent $a$ is linear in $c$), but the other group element part 
is uniquely determined independently of $c$. 
This means that, in the presence of nonzero central charge, the 
conformal transformation operator $U_f$ has an ambiguity of overall 
factor depending on which parameterization form is adopted in defining 
$U_f$. This is so because unit `operator' 1 is also one of the generators 
of the extended Virasoro algebra with central charge. 

The same problem of multiplicative c-number factor arises also in the 
composition law of two group elements. Whatever parameterization convention 
is adopted for $U_f$ and fixed, the multiplication of two elements 
$U_f$ and $U_g$ yields $U_{f\circ g}$ of composite mapping $f\circ g$
only up to a constant $e^a$:
\begin{equation}
U_f\cdot U_g = e^a\, U_{f\circ g}\,.
\end{equation}
Again the constant $a$ is linear in the central charge $c$, 
(and has complicated dependence both on the mappings $f,\ g$ and the 
parameterization convention). So {\it the naive composition law is 
violated unless the conformal anomaly is zero}. This is the crucial 
property which gives the reason why the GGRT holds only in the critical 
dimension. 

The characteristic feature of the normal ordered form 
(\ref{eq:Ufstandard}) for $U_f$ is that it manifestly satisfies 
\begin{equation}
\bra{3}U_f\ket{0}=
\bra{0}U_f\ket{3}=1
\qquad \bigl(\bra{3}\equiv\bra{0}c_\m1c_0c_1,\ \ 
\ket{3}\equiv c_\m1c_0c_1\ket{0} \bigr)
\label{eq:property}
\end{equation}
even when $c\not=0$. (Recall that the SL(2;C) invariant vacuum is normalized
by the condition 
$\bra{0}c_\m1c_0c_1\ket{0}=\left<3|0\right>=\left<0|3\right>=1$.) 
This property follows because:
the SL(2;C) invariant vacuum, either $\ket{0}$ or $\bra{0}$, is literally 
invariant under SL(2;C) transformation 
$\exp\bigl(\sum_{k=0,\pm1} v_{-k}L_k\bigr)$, the ghost-number 3 ket state 
$\ket{3}\equiv c_\m1c_0c_1\ket{0}$ is invariant under 
$\exp\bigl(\sum_{m\geq2} v_{-m}L_m\bigr)$ since 
$L_m\ket{3}=L_mc_\m1c_0c_1\ket{0}=0$ for $m\geq1$, and similarly, 
$\bra{3}$ is invariant under $\exp\bigl(\sum_{n\geq2} v_{n}L_{-n}\bigr)$
since $\bra{3}L_{-n}=\bra{0}c_\m1c_0c_1L_{-n}=0$ for $n\geq1$.

Now we have the following simple proposition, which was essentially
stated and used in LPP already:\cite{rf:LPP2}

\medskip
{\bf Proposition 1}\quad {\it Let $f(z)$ be a conformal mapping 
satisfying $f(0)=0$. If $f(z)$ is analytic and invertible in a 
neighborhood of $z=0$, then the corresponding operator $U_f$ defined by
Eqs.~(\ref{eq:Uf}) and (\ref{eq:OpTrf}), leaves the SL(2;C) ket vacuum 
inert up to a multiplicative constant:
\begin{equation}
U_f\ket{0} = e^a\,\ket{0}.
\end{equation}
If $U_f$ is taken to be of the normal ordered form (\ref{eq:Ufstandard}), 
or the conformal anomaly is absent, this constant $e^a$ equals} 1.
\smallskip

Proof) \ \ The SL(2;C) ket vacuum $\ket{0}$ is characterized by the 
property that $\phi(z)\ket{0}$ remains regular as $z\rightarrow0$ for any primary 
fields $\phi(z)$. This property implies for a primary field of dimension $d$,
$\phi(z)=\sum_{n}\phi_n z^{-n-d}$, that  
\begin{equation}
\lim_{z\rightarrow0}\phi(z)\ket{0}={\rm regular} \qquad 
\Leftrightarrow \qquad \phi_n\ket{0}=0 \quad {\rm for}\ \ n\geq1-d.
\label{eq:vacuum}
\end{equation}
So, if we can show that 
\begin{equation}
\lim_{z\rightarrow0}f[\phi(z)]\ket{0}={\rm regular},
\label{eq:regularity}
\end{equation}
for any primary fields $\phi$, then since 
$f[\phi(z)]=\sum_{n}f[\phi_n] z^{-n-d}$, we can deduce 
\begin{equation}
f[\phi_n]\ket{0}=U_f\phi_nU_f^\m1\ket{0}=0  
\quad \rightarrow\quad \phi_nU_f^\m1\ket{0}=0 \quad {\rm for}\ \ n\geq1-d
\end{equation}
This already implies that 
$U_f^\m1\ket{0}$ is proportional to the vacuum $\ket{0}$, 
$U_f^\m1\ket{0}=e^{-a}\,\ket{0}$, or equivalently, 
$U_f\ket{0} = e^a\, \ket{0}$
with some constant $a$. But, acting $\bra{3}$ to 
this relation from the left and using the normalization condition 
$\langle3|0\rangle=1$, we have
\begin{equation}
\bra{3}U_f\ket{0}=e^a \,.
\end{equation}
If $U_f$ is of the normal ordered form (\ref{eq:Ufstandard}), 
or the conformal anomaly is absent, 
the left-hand side is 1 by Eq.~(\ref{eq:property}), and $e^a=1$ follows.

Thus we have now only to prove Eq.~(\ref{eq:regularity}). The mapped field 
$f[\phi(z)]$ is explicitly given by Eq.~(\ref{eq:OpTrf}) for primary fields 
and so is expanded as 
\begin{equation}
f[\phi(z)] = \left(f'(z)\right)^d\phi\bigl(f(z)\bigr)
=\sum_{n}\phi_n\cdot \left(f(z)\right)^{-n-d}\left(f'(z)\right)^d\,.
\label{eq:expand}
\end{equation}
By the assumption of analyticity of $f(z)$ around the 
origin and $f(0)=0$, $f(z)$ behaves as $f(z)= f_1z + O(z^2)$. Moreover,
$f'(0)\not=0$ by the assumption of invertibility of $f(z)$ 
around $z=0$, and hence $f_1\not=0$. So, clearly, 
singular terms 
in the expansion 
(\ref{eq:expand}) as $z\rightarrow0$ are only those with $-n-d\leq-1$, but 
they all vanish on the vacuum $\ket{0}$ by Eq.~(\ref{eq:vacuum}), 
$\phi_n\ket{0}=0$ for $n\geq1-d$. Thus the condition 
(\ref{eq:regularity}) actually holds. \ q.e.d.

We can rewrite $U_f$ always into the normal ordered form up to 
a multiplicative constant. Then, let the general operator $U_f$ of the 
normal ordered form (\ref{eq:Ufstandard}) act on the ket vacuum:
\begin{eqnarray}
U_f\ket{0} &=& \exp\bigl(\sum_{n\geq2} v_{n}L_{-n}\bigr)
\exp\bigl(\sum_{k=0,\pm1} v_{-k}L_k\bigr) 
\exp\bigl(\sum_{m\geq2} v_{-m}L_m\bigr)\ket{0} \nn
&=& 
\exp\bigl(\sum_{n\geq2} v_{n}L_{-n}\bigr)\ket{0}
\end{eqnarray}
where use has been made of $L_n\ket{0}=0$ for $n\geq-1$. So, if this 
equals $\ket{0}$, we must have $v_n=0$ for $n\geq2$. This implies that 
$U_f$ has actually a simple form
\begin{equation}
U_f = \exp\bigl(\sum_{k=0,\pm1} v_{-k}L_k\bigr) 
\exp\bigl(\sum_{m\geq2} v_{-m}L_m\bigr)
\label{eq:simple}
\end{equation}
for such $f(z)$ analytic and invertible around the origin. 
In the case of infinitesimal transformation, $f(z) = z + \delta z$, this is 
an expected result from the beginning, since $L_{-n}$ is a generator of 
$\delta z=z^{-n+1}$ which is singular at $z=0$ for $n\geq2$.


In a similar manner, one can prove that the bra vacuum $\bra{0}$ 
remains intact under conformal transformations 
which have the same properties around the point at infinity. 

\medskip
{\bf Proposition 2}\quad {\it Let $f(z)$ be a conformal mapping 
satisfying the same conditions as in Proposition 1. 
Then the mapping $g \equiv I\circ f\circ I$ satisfies, schematically writing, 
$g(\infty)=\infty$ and is analytic and invertible in a neighborhood of 
$z=\infty$. The corresponding operator $U_g=U_{I\circ f\circ I}$ leaves the 
SL(2;C) bra vacuum 
inert up to a multiplicative constant:
\begin{equation}
\bra{0}U_g = e^a\,\bra{0}.
\end{equation}
If $U_g$ is taken to be of the normal ordered form (\ref{eq:Ufstandard}), 
or the conformal anomaly is absent, this constant $e^a$ equals} 1.
\medskip

Alternatively, this could also be proved as follows if we use the form 
(\ref{eq:simple}) for $U_f$ and the transformation property of $L_n$ 
under inversion, $I[L_n] = (-1)^nL_{-n}$:
\begin{eqnarray}
U_g &=& U_{I\circ f\circ I} = 
I\left[ \exp\bigl(\sum_{k=0,\pm1} v_{-k}L_k\bigr)
\exp\bigl(\sum_{m\geq2} v_{-m}L_m\bigr) \right] \nn
&=& \exp\bigl(\sum_{m\geq2} (-1)^mv_{-m}L_{-m}\bigr) 
\exp\bigl(\sum_{k=0,\pm1} (-1)^kv_{-k}L_{-k}\bigr)\,.
\end{eqnarray}
Then $\bra{0}U_g=\bra{0}$ would be clear since 
$\bra{0}L_{-n}=0$ for $n\geq-1$.

Without loss of generality, we can assume both $h_C$ and $h_D$ 
to map the origin of the unit disk to the origins in $M$ and $N$, 
respectively: 
\begin{equation}
h_C(w{=}0) = 0\,, \qquad h_D(w{=}0) = 0\,.
\end{equation}
This is achieved, if necessary, by performing SL(2;C) transformation on 
$M$ and $N$, since the CFT correlation functions are SL(2;C) invariant. 
Then, the mappings $h^\m1_C$ and $h^\m1_D$, as well as $h_C$ and $h_D$, 
satisfy the required properties of the propositions, and so we have 
\begin{equation}
U_{h_C^\m1}\ket{0}=\ket{0}\,, \qquad 
\bra{0}U_{I\circ h_D^\m1\circ I}=\bra{0}\,.
\label{eq:hchd}
\end{equation}
We shall frequently use a simple formula below which follows from 
Eq.~(\ref{eq:Uf}) immediately:
\begin{equation}
\langle \ f[\calO_1]f[\calO_2]\cdots\ \rangle 
= \bra{0}U_f\,\calO_1\calO_2\cdots\,U_f^\m1 \ket{0}\,.
\label{eq:formula}
\end{equation}

Now start the proof of GGRT, Eq.~(\ref{eq:GGRT}).
Using the LPP mapping relations (\ref{eq:mappingrel}) and 
the formula (\ref{eq:formula}), we first rewrite 
Eq.~(\ref{eq:GGRTvertex}) as 
\begin{eqnarray}
&&\v(\{B_j\},\{A_i\})
\prod_{i=1}^n \ket{A_i}_{A_i}
\prod_{j=1}^m \ket{B_j}_{B_j} \nn
&&\qquad = 
\bigl\langle \,\prod_{i=1}^n \hat h_{A_i}[\calO_{A_i}] \, 
\prod_{j=1}^m \hat h_{B_j}[\calO_{B_j}] \, \bigr\rangle_{M\infty N} \nn
&&\qquad = 
\bigl\langle \,\prod_{i=1}^n g\circ h^\m1_C\circ h_{A_i}[\calO_{A_i}] \, 
\prod_{j=1}^m g\circ I\circ h_D^\m1\circ h_{B_j}[\calO_{B_j}] 
\, \bigr\rangle_{M\infty N} \nn
&&\qquad = 
\bra{0}U_g \,\prod_{i=1}^n h^\m1_C\circ h_{A_i}[\calO_{A_i}] \, 
\prod_{j=1}^m I\circ h_D^\m1\circ h_{B_j}[\calO_{B_j}] 
\, U_g^\m1\ket{0}
\,.
\end{eqnarray}
Inserting the completeness relation (\ref{eq:completeness}) 
in the middle here, and applying the formula (\ref{eq:formula}) for 
$f=h_C^\m1$ and $I\circ h_D^\m1\circ I$, we have 
\begin{eqnarray}
&&= \sum_\alpha 
\bra{0}U_g \,\prod_{i=1}^n h^\m1_C\circ h_{A_i}[\calO_{A_i}] 
\,\calO_\alpha\ket{0} \bra{0}I[\calO_{\tilde\alpha}]\,
\prod_{j=1}^m I\circ h_D^\m1\circ h_{B_j}[\calO_{B_j}] 
\, U_g^\m1\ket{0}\nn
&&= \sum_\alpha 
\bra{0}U_gU_{h_C^\m1} \,\prod_{i=1}^n h_{A_i}[\calO_{A_i}] \,
h_C[\calO_\alpha]\,U^\m1_{h_C^\m1}\ket{0} \nn
&&\qquad \quad \times 
\bra{0}U_{I\circ h_D^\m1\circ I}\,I\circ h_D[\calO_{\tilde\alpha}]\,
\prod_{j=1}^m I\circ h_{B_j}[\calO_{B_j}] 
\, U^\m1_{I\circ h_D^\m1\circ I}U_g^\m1\ket{0}\,.
\end{eqnarray}
But, by Eq.~(\ref{eq:hchd}), we can use $U^\m1_{h_C^\m1}\ket{0}=\ket{0}$
and $\bra{0}U_{I\circ h_D^\m1\circ I}=\bra{0}$ to get
\begin{eqnarray}
&&= \sum_\alpha 
\bra{0}U_gU_{h_C^\m1} \,\prod_{i=1}^n h_{A_i}[\calO_{A_i}] \,
h_C[\calO_\alpha]\,\ket{0} \nn
&&\qquad \quad \times 
\bra{0}\,I\circ h_D[\calO_{\tilde\alpha}]\,
\prod_{j=1}^m I\circ h_{B_j}[\calO_{B_j}] 
\, U^\m1_{I\circ h_D^\m1\circ I}U_g^\m1\ket{0}\,.
\end{eqnarray}
Now suppose that the relations 
\begin{equation}
\bra{0}U_gU_{h_C^\m1}=\bra{0} \,, \qquad 
U^\m1_{I\circ h_D^\m1\circ I}U_g^\m1\ket{0}= \ket{0}\,,
\label{eq:g}
\end{equation}
hold, then we obtain
\begin{eqnarray}
&&= \sum_\alpha \,
\bra{0}\,\prod_{i=1}^n h_{A_i}[\calO_{A_i}] \,
h_C[\calO_\alpha]\,\ket{0}
\cdot \bra{0}\,I\circ h_D[\calO_{\tilde\alpha}]\,
\prod_{j=1}^m I\circ h_{B_j}[\calO_{B_j}] 
\, \ket{0}\nn
&&= \sum_\alpha \,
\bigl\langle \,\prod_{i=1}^n h_{A_i}[\calO_{A_i}] \,
h_C[\calO_\alpha]\,\bigr\rangle
\cdot \bigl\langle \,I\circ h_D[\calO_{\tilde\alpha}]\,
\prod_{j=1}^m I\circ h_{B_j}[\calO_{B_j}] 
\, \bigr\rangle\nn
&&= \sum_\alpha \,
\bigl\langle \,\prod_{i=1}^n h_{A_i}[\calO_{A_i}] \,
h_C[\calO_\alpha]\,\bigr\rangle
\cdot \bigl\langle \,h_D[\calO_{\tilde\alpha}]\,
\prod_{j=1}^m h_{B_j}[\calO_{B_j}] 
\, \bigr\rangle\ ,
\end{eqnarray}
where use has been made of the invariance of CFT correlation functions 
under inversion, i.e., $\VEV{I[\calO]}=\VEV{\calO}$,\footnote{Incidentally, 
this invariance leads to an identity 
$\left<A|B\right>=(-1)^{\abs{A}\abs{B}}\left<B|A\right>$, 
since $\left<A|B\right>=\VEV{I[\calO_A]\calO_B}=\VEV{\calO_AI[\calO_B]}
=(-1)^{\abs{A}\abs{B}}\VEV{I[\calO_B]\calO_A}
=(-1)^{\abs{A}\abs{B}}\left<B|A\right>$. This implies the symmetry 
property of the reflector, $\braR(12)=\braR(21)$.}
But the last expression is just identical with the definition of the 
vertices $\v(C,\{A_i\})$ and 
$\v(D,\{B_j\})$ in Eqs.~(\ref{eq:vertexA}) and (\ref{eq:vertexB}).
So, we obtain
\begin{eqnarray}
&=&\sum_\alpha \,\v(C,\{A_i\}) \prod_{i=1}^n \ket{A_i}_{A_i}
\ket{\alpha}_C \cdot
\v(D,\{B_j\}) \ket{\tilde\alpha}_D\prod_{j=1}^m \ket{B_j}_{B_j} \nn
&=&
\sum_\alpha \,\v(D,\{B_j\}) \v(C,\{A_i\}) \prod_{i=1}^n \ket{A_i}_{A_i}
\ket{\alpha}_C \,\ket{\tilde\alpha}_D\prod_{j=1}^m \ket{B_j}_{B_j} \nn
&=&
\v(D,\{B_j\}) \v(C,\{A_i\}) \prod_{i=1}^n \ket{A_i}_{A_i}
\ketR(CD)\,\prod_{j=1}^m \ket{B_j}_{B_j} \nn
&=&
\epsilon(A)\, \v(D,\{B_j\}) \v(C,\{A_i\}) \ketR(CD)\,
\prod_{i=1}^n \ket{A_i}_{A_i}\,\prod_{j=1}^m \ket{B_j}_{B_j}\,,
\hspace{1em}
\end{eqnarray}
where we have used the expression (\ref{eq:reflector}) for the 
reflector $\ketR(CD)$, and $\epsilon(A)$ is the sign factor defined 
in Eq.~(\ref{eq:sign})
which has appeared since we have changed the order of 
$\prod_{i=1}^n \ket{A_i}_{A_i}$ and 
the reflector $\ketR(12)$, the latter of which is Grassmann 
{\it odd}.\footnote{The Grassmann oddness comes from the fact
$\bra{0}{c_\m1 c_0 c_1}\ket{0}=1$ after all, and hence $\calO_\alpha$ 
and $\calO_{\tilde\alpha}$ must have opposite statistics in order to have 
non-vanishing inner product $\left<\tilde\alpha|\alpha\right>=1$. 
For the case of closed string, however, the reflector $\ketR(12)$ is 
Grassmann {\it even} since it is a product of two `open string' 
reflectors corresponding to the holomorphic and anti-holomorphic modes.} 
We thus obtain the desired identity:
\begin{equation}
\v(\{B_j\},\{A_i\}) = \epsilon(A) \,
\v(D,\{B_j\}) \v(C,\{A_i\}) \ketR(CD)\,.
\end{equation}

Now it is, therefore, sufficient to prove the relations (\ref{eq:g}). 
Recall that $g$ is a mapping which smoothes out the branch cuts 
generated in the plane $h_C^\m1(M)\infty\,I{\circ}h^\m1_D(N)$. So, although 
the mappings $g$ as well as $h_C^\m1$ and $h_D^\m1$ are singular, the 
composite mapping $g\circ h_C^\m1$ from $M$ to $M\infty N$, is analytic and 
invertible outside the image of the unit disk, $h_C(\abs{w}<1)$, in the 
$M$ plane, and so is the mapping $g\circ I\circ h_D^\m1$ from $N$ 
to $M\infty N$, outside the image of the unit disk, $h_D(\abs{w}<1)$, in the 
$N$ plane. Again, using the freedom of SL(2;C) transformation in the 
$M\infty N$ plane, we can make the mapping $g$ to satisfy 
\begin{equation}
g\circ h_C^\m1(\infty)=\infty\,, \qquad 
g\circ I\circ h_D^\m1\circ I(0)=0\,.
\end{equation}
Then, the mappings $g\circ h_C^\m1$ and $g\circ I\circ h_D^\m1\circ I$ 
are analytic and invertible in a neighborhood of $z=\infty$ and $z=0$, 
respectively, and so we can apply the propositions 2 and 1 to obtain
\begin{equation}
\bra{0}U_{g\circ h_C^\m1}=\bra{0}\,, \qquad 
U_{g\circ I\circ h_D^\m1\circ I}\ket{0}=\ket{0}\,. 
\end{equation}  
These give just the desired relations (\ref{eq:g}), if 
\begin{equation}
U_{g\circ h_C^\m1} = U_gU_{h_C^\m1}\,, \qquad 
U_{g\circ I\circ h_D^\m1\circ I}=U_gU_{I\circ h_D^\m1\circ I}\,.
\end{equation}
These actually hold with {\it weight one} when {\it the conformal 
anomaly is zero}. Other than in critical dimension, there appears very 
non-trivial multiplicative c-number factor by which
Eq.~(\ref{eq:GGRT}) is violated. This finishes the proof of GGRT.

\section{GGRT at one loop}

Next we prove here a generalized version of GGRT at one loop level; 
that is, the theorem for the fused vertex with double contractions by 
two reflectors, which read something like
\begin{equation}
\v(D,F,\{B_j\}) \v(C,\{A_k\},E) \ketR(CD)\ketR(EF) = 
\epsilon_L \,\bra{v_L(\{B_j\},\{A_k\})}\,.
\label{eq:GGRTloop}
\end{equation}
Here the suffix $L$ denotes the quantities at 1-loop level whose 
more precise definitions will be given in the course of the proof. 

Using the GGRT at tree level, we can first rewrite the 
left-hand side into
\begin{eqnarray}
&&\v(D,F,\{B_j\}) \v(C,\{A_k\},E) \ketR(CD)\ketR(EF) \nn
&&\qquad \qquad = 
\,\v(F,\{\Phi_i\},E) \epsilon(E+A) \ketR(EF)
\label{eq:loop1}
\end{eqnarray}
with abbreviation $\{\Phi_i\}$ denoting the combined set of states 
$\{B_j\}$ and $\{A_k\}$, where $\v(F,\{\Phi_i\},E)$ is a 
tree level LPP vertex obtained by the fusion of the two vertices by 
a single contraction $\ketR(CD)$, and $\epsilon(E+A)$ is 
the operator sign factor $(\m1)^{\abs{E}+\sum_k\abs{A_k}}$ according to 
Eq.~(\ref{eq:sign}). 
Since this vertex $\v(F,\{\Phi_i\},E)$ is 
a tree level one, it corresponds to a plane, which we call $M$, and 
there are mappings $h_F,\ h_E$ and $h_{\Phi_i}$ which map the unit disks 
of the strings $F$, $E$ and $\Phi_i$ into the plane $M$, analytic and 
invertible inside each unit disk, respectively:
\begin{eqnarray}
&&\v(F,\{\Phi_i\},E) \epsilon(E+ A)\ketR(EF)
\prod_i \ket{\calO_{\Phi_i}}_{\Phi_i} \nn
&& \qquad \quad = \sum_\alpha \,(\m1)^{\abs{\alpha}+\sum_k\abs{A_k}}
\v(F,\{\Phi_i\},E) \ket{\alpha}_E\ket{\tilde\alpha}_F\,
\prod_i \ket{\calO_{\Phi_i}}_{\Phi_i} \nn
&& \qquad \quad = (\m1)^{\sum_k\abs{A_k}}\sum_\alpha \,(\m1)^{\abs{\alpha}}
\bigl\langle \,h_E[\calO_\alpha]\,h_F[\calO_{\tilde\alpha}]\,
\prod_{i=1}^n h_{\Phi_i}[\calO_{\Phi_i}] \,\bigr\rangle_M \,.
\label{eq:loop2}
\end{eqnarray}
\begin{wrapfigure}[11]{r}{\halftext}
   \epsfxsize= 6 cm   
   \centerline{\epsfbox{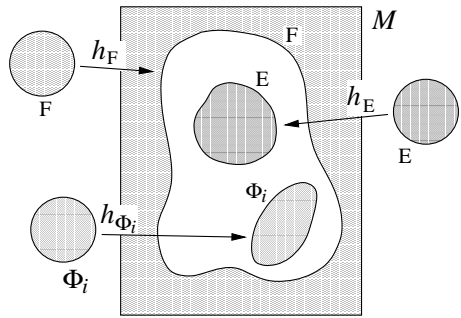}}
 \vspace{2pt}
 \caption{Mappings defining $\v(F,\{\Phi_i\},E)$.}
 \label{fig:2}
\end{wrapfigure}
Using the freedom of 
SL(2;C), we can assume without loss of generality that
\begin{equation}
h_E(w{=}0)=0\,, \quad 
h_F(w{=}0)=\infty\,. 
\end{equation}
These mappings are schematically shown in Fig.~\ref{fig:2}. 
The further contraction by $\ketR(EF)$ in Eq.~(\ref{eq:loop1}), or 
summation over $\alpha$ in Eq.~(\ref{eq:loop2}), 
corresponds to the gluing of the two boundaries 
$h_E(\abs{w}{=}1)$ and $h_F(\abs{w}{=}1)$ in this plane $M$, which makes
the plane a torus, which we call $M8$. The torus $M8$ can be 
represented by a complex plane with identification 
\begin{equation}
z \ \sim\  q^2z\,, \qquad \exists q\equiv e^{i\pi\tau}
\end{equation}
This means that there is a smooth mapping of $M$ into the torus plane $M8$, 
and there are mappings $\hat h_{\Phi_i}$ of the unit disks to $M8$ which are 
analytic and invertible inside each unit circle.

\begin{figure}[tb]
   \epsfysize= 13 cm
   \centerline{\epsfbox{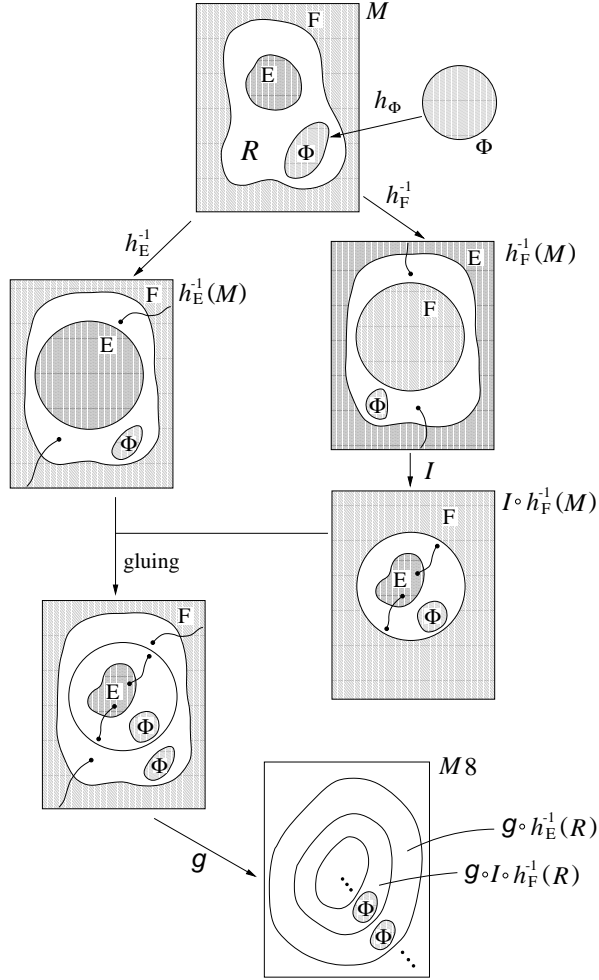}}
 \caption{Gluing and resmoothing procedure to obtain the loop vertex 
          $\bra{v_L(\{\Phi_i\})}$.}
 \label{fig:3}
\end{figure}

In string field theory, this mapping can be decomposed into the 
following steps in a very similar manner to LPP at the tree level case. 
As shown in Fig.~\ref{fig:3}, 
the complex plane $M$ is mapped 
in two ways, one by $h_E^\m1$ and the other by $I\circ h_F^\m1$, so that 
the exterior region of string $E$ in $M$ is mapped to the region outside 
a unit circle and the exterior region of string $F$ in $M$ to the 
inside of a unit circle, respectively. Then the region outside the 
unit circle in the plane $h_E^\m1(M)$ and the region inside the unit 
circle in the plane $I\circ h^\m1_F(M)$ are glued smoothly as they stand. 
But, again, the glued surface generally possesses branch cuts  
unless the mappings $h_C$ and $h_D$ are $SL(2;C)$ transformations. 
Since we know the glued surface is a covering space of a torus in any case, 
there exists a mapping $g$ which carries the surface into the torus 
plane $M8$ (the plane with identification $z\sim q^2z$), smoothing out the 
branch cuts. Therefore the conformal mappings $\hat h_{\Phi_i}$ 
of the strings $\{\Phi_i\}$ into the torus plane $M8$,
mentioned above, are identified with
\begin{equation}
\hat h_{\Phi_i} = g\circ h_E^\m1\circ h_{\Phi_i}\,.
\label{eq:mappingrelloop}
\end{equation}
But, in this loop case, the mappings via the other route, 
$g\circ I\circ h_F^\m1\circ h_{\Phi_i}$, should equally be good mappings. 
Indeed, the whole region $R$ in $M$ outside the both images of the unit 
disks of strings $E$ and $F$, is mapped to the two adjacent regions 
in $M8$ displaced with a period $q^2$ if we follow the two routes of 
mappings, $g\circ h_E^\m1$ and $g\circ I\circ h_F^\m1$. This can be 
easily seen by inspecting Fig.~\ref{fig:3}. That is, we have 
the following equation for $\forall z\in R$ in $M$ 
\begin{equation}
g\circ I\circ h_F^\m1(z) = q^2 \times 
g\circ h_E^\m1(z)\,.
\label{eq:cnumkeyrel}
\end{equation}
This is a key relation in this one-loop case. This is purely a 
c-number relation between the two conformal mappings. The corresponding 
operator relation of course reads
\begin{equation}
U_{g\circ I\circ h_F^\m1} = U_{q^2\circ g\circ h_E^\m1}\,,
\end{equation}
where $q^2$ as a mapping denotes $q^2(z)=q^2\cdot z$. 
We know that the operator representation of this Weyl transformation 
$q^2$ is given by $U_{q^2}= q^{2L_0}$. Therefore, 
if the conformal anomaly is zero, we can use the composition law 
for the group elements freely to obtain
\begin{eqnarray}
&&U_{g\circ I\circ h_F^\m1} = q^{2L_0}\,U_{g\circ h_E^\m1} \nn
&&\ \rightarrow \quad U_{h_F\circ I}^\m1
= U_{I\circ h_F^\m1} = U_g^\m1\,q^{2L_0}\,U_{g\circ h_E^\m1}\,.
\label{eq:keyrel}
\end{eqnarray}
If $c\not=0$, there will appear  non-trivial multiplicative 
c-number factors in these equations. 

With these equations at hand, we can now prove the GGRT at one-loop as 
follows. We rewrite Eq.~(\ref{eq:loop2}), aside from the fixed 
sign factor $(\m1)^{\sum_k\abs{ A_k}}$, into
\begin{eqnarray}
&&\sum_\alpha \,(\m1)^{\abs{\alpha}}\,
\bigl\langle \,h_E[\calO_\alpha]\,h_F[\calO_{\tilde\alpha}]\,
\prod_{i=1}^n h_{\Phi_i}[\calO_{\Phi_i}] \,\bigr\rangle_M \nn
&&\qquad = \sum_\alpha \,(\m1)^{\abs{\alpha}}\,\epsilon' \,
\bigl\langle \,h_F[\calO_{\tilde\alpha}]\,
\prod_{i=1}^n h_{\Phi_i}[\calO_{\Phi_i}] \,
h_E[\calO_\alpha]\,\bigr\rangle_M \nn
&&\qquad = \sum_\alpha \,(\m1)^{\abs{\alpha}}\,\epsilon' \,
\bra{0}\,U_{h_F\circ I}\,I[\calO_{\tilde\alpha}]\,U^\m1_{h_F\circ I}\,
\prod_{i=1}^n h_{\Phi_i}[\calO_{\Phi_i}] \,h_E[\calO_\alpha]\,\ket{0}
\hspace{1em}
\label{eq:loop3}
\end{eqnarray}
with a new sign factor 
$\epsilon'=(\m1)^{\abs{\alpha}(\abs{\tilde\alpha}+\sum_i\abs{\Phi_i})}$.
Fortunately, however, this sign factor is 1 since 
$\abs{\alpha}+\abs{\tilde\alpha}+\sum_i\abs{\Phi_i}=1$ 
mod 2 in order for the correlation function in $M\sim S^2$ to be non-zero, 
implying $\abs{\alpha}(\abs{\tilde\alpha}+\sum_i\abs{\Phi_i})
=\abs{\alpha}(\abs{\alpha}+1)=0$ mod 2. 
Here, since $h_F\circ I(\infty)=\infty$ and $h_F\circ I$ is analytic and 
invertible in a neighborhood of $z=\infty$, we can use
$\bra{0}U_{h_F\circ I}=\bra{0}$ by Proposition 2. 
Moreover we can use the above key relation (\ref{eq:keyrel})
when $c=0$. Then we further proceed as follows:
\begin{eqnarray}
&=& \sum_\alpha \,(\m1)^{\abs{\alpha}}\,
\bra{0}I[\calO_{\tilde\alpha}] \,U_g^\m1\,q^{2L_0}\,U_{g\circ h_E^\m1}\,
\prod_{i=1}^n h_{\Phi_i}[\calO_{\Phi_i}] \,h_E[\calO_\alpha]\,\ket{0} \nn
&=& \sum_\alpha \,(\m1)^{\abs{\alpha}}\,
\bra{0}I[\calO_{\tilde\alpha}] \,U_g^\m1\,q^{2L_0}\,U_{g\circ h_E^\m1}\,
U_{h_E}\prod_{i=1}^n h_E^\m1\circ h_{\Phi_i}[\calO_{\Phi_i}] \,
\calO_\alpha\,U_{h_E}^\m1\ket{0} \nn
&=& \sum_\alpha \,(\m1)^{\abs{\alpha}}\,
\bra{0}I[\calO_{\tilde\alpha}] \,U_g^\m1\,q^{2L_0}\,U_g\,
\prod_{i=1}^n h_E^\m1\circ h_{\Phi_i}[\calO_{\Phi_i}] \,
\calO_\alpha\ket{0} \nn
&=& \Tr \bigl[\,U_g^\m1\,q^{2L_0}\,U_g\,
\prod_{i=1}^n h_E^\m1\circ h_{\Phi_i}[\calO_{\Phi_i}] \,\bigr]\,.
\end{eqnarray}
Here, in going to the third line, we have used 
$U_{h_E}^\m1\ket{0}=\ket{0}$ by Proposition 1 and $U_{g\circ 
h_E^\m1}U_{h_E}=U_g$ which holds again when $c=0$. To the last 
expression we have used a {\it definition}
\begin{equation}
\Tr[\,\cdots\,] = \sum_\alpha \,(\m1)^{\abs{\alpha}}\,
\bra{0}I[\calO_{\tilde\alpha}] \,\cdots \,\calO_\alpha\ket{0} 
= \sum_\alpha \,(\m1)^{\abs{\alpha}}\,\bra{\tilde\alpha}\,\cdots 
\,\ket{\alpha}\,. 
\label{eq:supertrace}
\end{equation}
It may have sounded strange to call this a `definition': of course this 
$\Tr$ is just the usual trace for the bosonic mode sector and 
the usual super trace for the ghost non-zero mode sector.  However the 
trace operation for the ghost zero-mode sector is not so self-evident (as 
will be explained later) and this gives the definition for it. The usual
cyclic identity for bosonic operators also holds for this trace.

It is now immediate to rewrite the last trace expression into the 
final form:
\begin{eqnarray}
&=& \Tr \bigl[\,U_g^\m1\,q^{2L_0}\,U_g\,
\prod_{i=1}^n h_E^\m1\circ h_{\Phi_i}[\calO_{\Phi_i}] \,\bigr]
= \Tr \bigl[\,q^{2L_0}\,U_g\,
\prod_{i=1}^n h_E^\m1\circ h_{\Phi_i}[\calO_{\Phi_i}] \,U_g^\m1\,\bigr] \nn
&=& \Tr \bigl[\,q^{2L_0}\,
\prod_{i=1}^n g\circ h_E^\m1\circ h_{\Phi_i}[\calO_{\Phi_i}]\,\bigr] 
= \Tr \bigl[\,q^{2L_0}\,
\prod_{i=1}^n \hat h_{\Phi_i}[\calO_{\Phi_i}]\,\bigr]\,,
\label{eq:loopfinal}
\end{eqnarray}
where the mapping relation (\ref{eq:mappingrelloop}) has been used. 

Thus, if we define the CFT correlation function on the torus by 
\begin{equation}
\left< \, \calO_1\,\calO_2\,\cdots\, \right>_{{\rm torus}\ \tau}=
\Tr[ \,q^{2L_0}\,\calO_1\,\calO_2\,\cdots\, ]\,,
\label{eq:loopgreenfn}
\end{equation}
and the LPP vertex at one-loop level by 
\begin{equation}
\bra{v_L(\{\Phi_i\};\tau)}\prod_i\ket{\Phi_i}_{\Phi_i} = 
\bigl\langle \, \prod_i\hat h_{\Phi_i}[\Phi_i]\, \big\rangle_{{\rm torus}
\ \tau}\,,
\end{equation}
then, what we have proved is summarized in the following GGRT at one-loop, 
by Eqs.~(\ref{eq:loop1}), (\ref{eq:loop2}),
(\ref{eq:loop3}) and  (\ref{eq:loopfinal}).

\medskip
{\bf Theorem}\ \ {\it When the conformal anomaly is zero, the fused vertex 
obtained by twice contractions by two reflectors, equals the LPP vertex 
at one-loop level up to sign}:
\begin{equation}
\v(D,F,\{ B_j\}) \v(C,\{ A_k\},E) \ketR(CD)\ketR(EF) = 
\epsilon( A) \,\bra{v_L(\{ B_j\},\{ A_k\};\tau)}\,,
\label{eq:GGRTloopfinal}
\end{equation}
{\it where the operator sign factor $\epsilon( A)$ is given by} 
$\epsilon( A) = (\m1)^{\sum_k\abs{ A_k}}$.
\medskip

A few remarks may be in order here:

If we define correlation functions on a torus for a system possessing 
non-zero central charge $c$, it is known better to replace 
$q^{2L_0}$ in Eq.~(\ref{eq:loopgreenfn}) with
$q^{2(L_0-c/24)}$:
\begin{equation}
\left< \, \calO_1\,\calO_2\,\cdots\, \right>_{{\rm torus}\ \tau}=
\Tr[ \,q^{2(L_0-c/24)}\,\calO_1\,\calO_2\,\cdots\, ]\,.
\label{eq:improved}
\end{equation}
This operator $L_0-c/24$ can be identified with $(L_0)_{\rm cylinder}$ 
on the cylinder ($\rho$-plane), and $-c/24$ comes from the Schwarzian 
derivative term in the anomalous transformation law of the energy 
momentum tensor under the coordinate change $\rho\rightarrow z=e^\rho$. 
Indeed, with this factor $(q^2)^{-c/24}$, the vacuum functional 
(partition function) becomes invariant under the modular transformation 
$\tau\rightarrow-1/\tau$ and the other correlation functions also turn 
to have more natural modular transformation properties. However note 
that this `improvement' does not cure at all the violation of the above 
GGRT when the net central charge is non-zero; the multiplicative 
c-number factor violating the theorem is a number which depends on all 
the details of the mappings and moduli parameters which cannot be 
cancelled by such a simple factor like $(q^2)^{-c/24}$. If the net 
central charge is zero, then one may of course calculate the correlation
functions for each sectors possessing central charges separately, by 
using the improved definition Eq.~(\ref{eq:improved}).

It is interesting that the super trace formula Eq.~(\ref{eq:supertrace})
appeared automatically in our derivation. It is by no means a priori 
clear how the trace should be defined for the 
ghost zero-mode sector, since it has a off-diagonal metric structure:
\begin{equation}
\left<2|1\right> =\left<1|2\right> = \bra{\Omega}c_0\ket{\Omega} = 1\,,
\end{equation}
where $\ket{1}\equiv c_1\ket{0}\equiv\ket{\Omega}$ and 
$\bra{1}\equiv\bra{0}c_\m1\equiv\bra{\Omega}$ are ket and bra Fock 
vacua, and $\ket{2}\equiv c_0\ket{\Omega}$ and 
$\bra{2}\equiv\bra{\Omega}c_0$. If we follow the definition 
(\ref{eq:supertrace}) of supertrace $\Tr$, then noting that 
$\abs{\alpha}= 1$ and 0 for $\ket{\alpha}= \ket{1}$ and $\ket{2}$, 
respectively, the trace in the ghost zero-mode sector, denoted by 
$\Tr_0$, is calculated as 
\begin{eqnarray}
\Tr_0[1] &=& (\m1)^1\bra{2}1\ket{1}+(\m1)^0\bra{1}1\ket{2} = -1 +1 =0,\nn
\Tr_0[c_0] &=& (\m1)^1\bra{2}c_0\ket{1}+(\m1)^0\bra{1}c_0\ket{2} = 0+0 =0,\nn
\Tr_0[b_0] &=& (\m1)^1\bra{2}b_0\ket{1}+(\m1)^0\bra{1}b_0\ket{2} = 0+0 =0,\nn
\Tr_0[c_0b_0] &=& (\m1)^1\bra{2}c_0b_0\ket{1}+(\m1)^0\bra{1}c_0b_0\ket{2} 
= 0 +1 = 1,
\end{eqnarray}
and $\Tr_0[b_0c_0] = \Tr_0[1-c_0b_0] = -1$, of course. 
These equations precisely show that the ghost correlation functions on 
the torus vanishes unless both $c_0$ and $b_0$ modes appear at least once, 
in conformity with the fact that there is one zero mode each for 
$c(z)$ and $b(z)$ in the torus case (that is, a conformal Killing vector
and a holomorphic quadratic differential, respectively).

It may be noted that the supertrace can also be rewritten into a form 
as given by Friedan et al,\cite{rf:FGST}
\begin{equation}
\Tr[\,\cdots\,] = -{\rm Tr}[\,(\m1)^{N_{\rm FP}} \,\cdots \,]
\end{equation}
in terms of the `usual' trace Tr (with understanding 
that ${\rm Tr}_0 [\calO] \equiv \bra{2}\calO\ket{1}+\bra{1}\calO\ket{2}$ 
in the zero mode sector) and the FP ghost number defined by
\begin{equation}
N_{\rm FP} = c_0b_0 + \sum_{n\geq1}(c_{-n}b_n - b_{-n}c_n) 
\end{equation}
which counts the ghost number from the Fock vacuum. 

\medskip 

The authors would like to thank K.~Suehiro and N.~Sasakura for helpful 
discussions.  T.~K.\ and T.~T.\ are supported in part by
the Grant-in-Aid for Scientific Research (\#10640261) and the
Grant-in-Aid (\#6844), respectively, from the Ministry of Education, 
Science, Sports and Culture.

\end{document}